\documentclass[secnumarabic,amssymb,amsmath, nobibnotes,aps,prl,reprint]{revtex4-1}

\usepackage{graphicx,subfigure,float,amssymb,amsmath,color}
\usepackage{mathdots}
\usepackage{xcolor}
\usepackage{bm}
\usepackage{csquotes}
\begin{document}


\newcommand{\Eqref}[1]{Eq.~(\ref{#1})}
\newcommand{\Eqsref}[2]{Eqs.~(\ref{#1}) and (\ref{#2})}
\newcommand{\refcite}[1]{Ref.~\cite{#1}}
\newcommand{\figref}[1]{Fig.~(\ref{#1})}
\newcommand{\Tref}[1]{Table ~(\ref{#1})}
\newcommand{\blue}[1]{\textcolor{blue}{#1}}
\newcommand{\red}[1]{\textcolor{red}{#1}}


\title{ Voltage fluctuations in ac biased superconducting transition edge sensors.}



\author{L. Gottardi} 
\email[]{Electronic mail: l.gottardi@sron.nl}
\affiliation{NWO-I/SRON Netherlands Institute for Space Research, Sorbonnelaan 2, 3584 CA Utrecht, The Netherlands} 
\author{M. de Wit}
\affiliation{NWO-I/SRON Netherlands Institute for Space Research, Sorbonnelaan 2, 3584 CA Utrecht, The Netherlands}         
\author{A. Kozorezov}
\affiliation{Department of Physics, Lancaster University, LA1 4YB, Lancaster, UK}                    
\author{E. Taralli}
\author{K. Nagayashi}
\affiliation{NWO-I/SRON Netherlands Institute for Space Research, Sorbonnelaan 2, 3584 CA Utrecht, The Netherlands} 




\date{\today}

\begin{abstract}
 We present a detailed analysis of the fundamental noise sources in  superconducting transition edge sensors (TESs), ac voltage biased at MHz frequencies and treated as superconducting weak-links. We have studied the noise in the resistive transition as a function of bath temperature  of several detectors with different normal resistances and geometries. We show that the \enquote*{excess} noise, typically observed  in the TES electrical bandwidth,  can be explained by the equilibrium Johnson noise of the quasiparticles generated within the  weak-link. The fluctuations at the Josephson frequency and higher harmonics contribute significantly to the measured voltage noise at the detector bandwidth through the non-linear response of the weak-link with a sinusoidal current-phase relation.   
\end{abstract}


\maketitle 



%
%

%

Superconducting transition-edge sensors (TESs) are very sensitive
thermometers used as microcalorimeters and bolometers in ground and space-borne low temperature instruments \cite{IrwinHilton,Ullom_2015}. 
TESs are,  typically, low impedance devices made of a thin superconducting bilayer, with a critical temperature $T_\mathrm{c}$.
They operate in the voltage  bias regime. Either a constant or an alternating bias voltage is used, depending on the multiplexing read-out scheme.  In the frequency division multiplexing (FDM) configuration considered here, the TES is ac voltage biased in the superconducting transition, by means of   high-{\it Q} {\it LC} resonators at frequencies from 1 up to 5 MHz. 
The underlying physics of TESs has been extensively studied in the past years and the TES response under ac and dc bias is fairly well described for TESs behaving as superconducting weak-links
\cite{Sadleir10,Smith13,Gottardi18,Gottardi_APL14}, or affected by the generation of phase-slip lines \cite{BennetPRB13}. 

\begin{figure}
\center
\includegraphics[width=7.5cm]{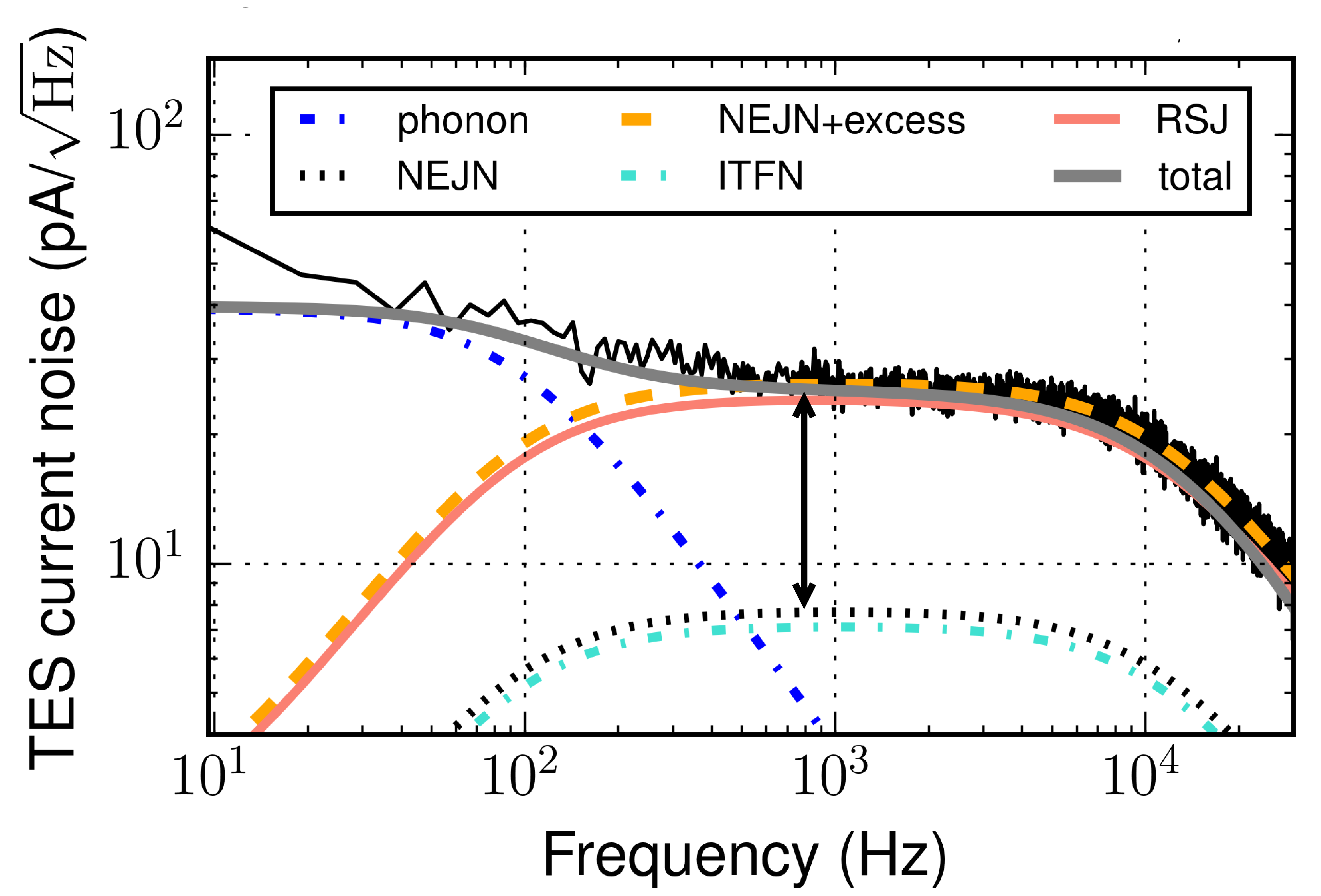}
\caption{\label{fig:schemeivnoise} Current noise spectrum for a TiAu $80\times 20 \, \mu \mathrm{m}^2$ TES biased at 2.6 MHz, $R/R_N=13\%$ and $T_{bath}=55\,\mathrm{mK}$. The  different lines show the  noise contributions discussed in the text. The vertical arrow indicates the measured excess noise with respect to the estimated NEJN.}
\end{figure}

 The basic theory of a  TES is extensively reported elsewhere \cite{IrwinHilton,Smith13,Ullom_2015}. The electrical and thermal equations for a TES are  solved exactly in the small signal regime with   resistance dependency on temperature $T$ and current $I$ linearly expanded to the first-order as  $R(I,T))=R_0+\alpha\frac{R_0}{T_0}\delta T+\beta\frac{R_0}{I_0}\delta I$. Here,  $\alpha=(T/R)(\partial R/\partial T)|_{I_0}$, $\beta=(I/R)(\partial R/\partial I)|_{T_0}$,  $\delta T=T-T_0$ and $\delta I =I-I_0$. The $\alpha$ and $\beta$ parameters can be measured experimentally at the operating point and are used to estimate the detector noise and sensitivity. Three fundamental contributions to the TES noise are generally identified. The first one is called {\it phonon noise}  from thermal fluctuations between the TES-absorber body and the heat bath, with power spectral density given by $S_{ph}=4k_BT^2G_{bath}((T_{bath}/T)^{n+2}+1)/2$, where $G_{bath}$ is the thermal conductance to the bath at a temperature $T_{bath}$ and $n$ is the exponent depending  on the nature of thermal processes involved. This noise is dominant at low frequencies in the detector thermal bandwidth typically below $\sim 200\mathrm{Hz}$.  The second contribution is the {\it Johnson-Nyquist noise} (JN) of the TES biased in the resistive transition. It has been modelled so far as a non-equilibrium Johnson noise (NEJN) with voltage spectral density $S_{V,nejn}=4k_BT\operatorname{Re}(Z_{tes})(1+2\beta)$ \cite{Irwin2006}. It is suppressed  at low frequency by the electro-thermal feedback \cite{Irwin1995} and becomes significant in the detector electrical band at kHz. The third noise term is the {\it internal thermal fluctuation noise} (ITFN), which is generated by thermal fluctuation between distributed heat capacities internal to the TES-absorber system. It has a power spectral density $S_{itfn}=4k_BT^2G_{tes}$, where $G_{tes}$ is the intrinsic thermal conductance of the system. The response of the detector to this noise source  is  identical to  the JN noise, which complicates the identification of the TES noise sources in the electrical bandwidth. The  ITFN can be derived from a proper characterization of the thermal circuit \cite{Hoevers00,Takei08,Kinn12,Maasilta12,Wake2019,*Wake2020}.
\noindent In \figref{fig:schemeivnoise}, we show  the typical current noise spectrum after demodulation of a MHz biased TES operating at $R/R_N=20\%$, where $R_N$ is the TES normal resistance. More details on how to derive the total current noise measured in a TES from the noise sources described above is given in the supplemental material \cite{SM}. 
So far, the total TES noise in the JN bandwidth has not been fully understood. The residual $M^2=(S_{V,data}-S_{V,model})/S_{V,model}$ is a convenient way to characterize the \enquote*{excess} noise. The common practice in the TES literature is to define $M^2$ with respect to the NEJN, i.e. $S_{V,model}=S_{V,nejn}$
 \cite{Ullom04,Takei08,Smith13,Wake2019,*Wake2020}.  

\noindent In this Letter, we study the Johnson-Nyquist (JN) noise in many ac biased TESs  with different electro-thermal properties. We show that the fluctuation-dissipation theorem generalized for  a non-linear system in thermal equilibrium explains well the observed noise and  that it is not necessary to introduce the formalism for a non-linear TES out-of-equilibrium \cite{Irwin2006}. The observed spectral density of fluctuations of the TES voltage is in full agreement with  the expected equilibrium thermal noise of the quasi-particles.
We characterized, at different bias frequencies, many Ti/Au TES microcalorimeters, with critical temperature $T_\mathrm{c}\sim 90\, \mathrm{mK}$,  normal sheet resistance $R_{\square}=26\, \mathrm{m}\Omega/ \square$ and three different geometries (length $\times$ width): $80\times 40$, $80\times 20$ and $120\times 20\, \mu \mathrm{m}^2$, leading to different values of $R_N$, $\alpha$, $\beta$, and saturation power. In this way we can probe the noise model in different weak-link regimes \cite{deWit_2020}. More details on the devices and the read-out system are given in \cite{deWit_2020,SM}.  
To study the intrinsic  fluctuations inside the TES, we consider  the seminal work of Likharev and Semenov \cite{LikSem72} and later reviews \cite{KoganBook,Vyst74}. We consider the  RSJ model previously proposed to explain the excess noise in  dc-biased TES  \cite{Kozo12}, extend it to the ac bias case and provide experimental evidence of its validity. Moreover, as proposed in \cite{Lhotel2007,Wessel2019}, we compare the RSJ approximation with the generalized kinetic theory for fluctuations in superconductors derived by Kogan and Nagaev (KN) \cite{KogNag88} and show that the two models are consistent with each other over a wide range of bias conditions, detector $\beta$ values and for different TES geometries. In \cite{Wessel2019}, only the results for two dc biased TESs are reported and a relatively large discrepancy  between the two models is observed. Following the method described in this letter,  the detector noise is  accurately estimated since we measure directly the physical  quantities required by the RSJ and KN  models and no free parameters are left. 
\begin{figure*}[ht]
\center
\includegraphics[width=16.7 cm]{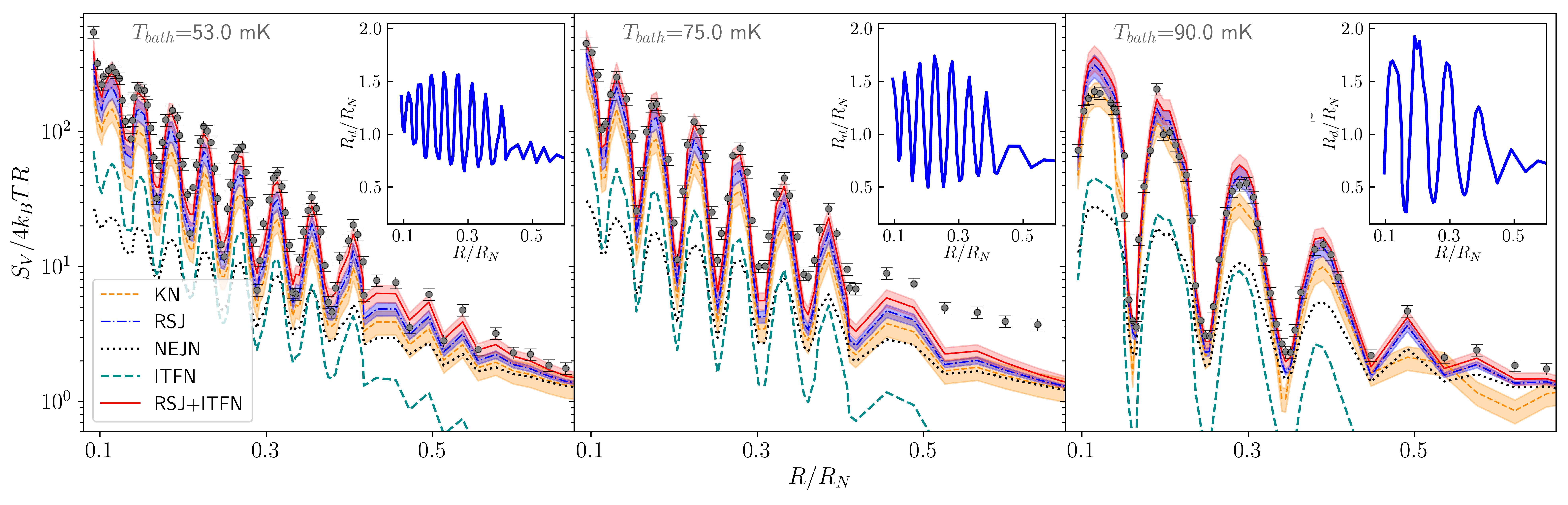}
\caption{\label{fig:SvvsRRn} Measured voltage noise of the $80\times 40 \, \mu\mathrm{m}$ TES as a function of bias point $R/R_N$.  From top to bottom: $T_{bath}=53.0,75.0$ and 90.0 mK. The blue dot-dashed  and the orange dashed lines are the prediction from  \Eqref{SvRSJ} and  \Eqref{SvKN}, respectively. The green dashed line is the estimated ITFN. The black dotted line is the NEJN. The red line is the sum in quadrature of the RSJ and ITFN noise.  Insets: the normalized TES dynamic resistance, $R_d=R(1+\beta)$ as a function of $R/R_N$.}
\end{figure*}  
A fundamental difference between a dc and ac biased TES is given by the evolution of the superconducting phase across the $SS'S$ or $SNS$ weak-link. Under dc bias, due to the ac Josephson effect, the phase $\varphi(t)$ increases linearly with time at a rate $\omega_J=2eV_{dc}/\hbar$ proportional to the bias voltage, where $\omega_J$ is the bias dependent Josephson frequency. Under ac bias, on the contrary,  $\varphi(t)$    oscillates around an equilibrium value $\varphi_0$ at a frequency $\omega_J=\omega_b$, out-of-phase with respect to the ac bias voltage $V(t)=V_{ac}cos(\omega_b t)$, and with a peak amplitude $\varphi_{pk}=2eV_{ac}/\hbar \omega_b$.
Despite the difference, when studying the voltage fluctuations within the weak-link context, both the dc and ac biased TES can be seen as a periodical non-stationary system with period $2\pi/\omega_J$. It is then convenient to represent the time dependent system as a combination of Fourier series over frequencies $\omega_m=\omega+m\omega_J$ and solve the problem of the weak-link dynamics as done by Zorin \cite{Zorin81,KoganBook}.
It is shown that the  voltage power spectral density of the  fluctuation $S_V(\omega)$ is the results of parametric conversion of fluctuations mixed  with the Josephson oscillations and its harmonics. The relation with the current power spectrum $S_I(\omega)$ is given by  $S_V(\omega)=\sum{Z^*_{mm'}(\omega)Z_{mm'}(\omega)S_I(\omega_m)}$. Here, $Z_{mm'}(\omega)$ is the impedance matrix of the RSJ, with the indices $m$ and $m'$ standing for the respective harmonics of the Josephson oscillation at $\omega_J$ \cite{LikSem72,KoganBook}. When measuring the noise at $\omega\ll\omega_J$, the contribution from the higher harmonics with $\left| m\right|>1$ was shown to be negligible \cite{Koch82}. 
The low frequency power spectral density of the voltage fluctuations, averaged over the period of oscillation $2\pi/\omega_J$ for an RSJ can be approximated
 \cite{Kozo12,AslLar1969,KoganBook,SM} to 
\begin{equation}\label{SvRSJ}
S_V(\omega)\simeq R_d^2\left[\frac{3}{2}\frac{R}{R_N}\left(1-\frac{1}{3}\left(\frac{R}{R_N}\right)^2\right)\right]S_I(\omega),
\end{equation}
\noindent where $R_d=\partial V/\partial I=R(1+\beta)$ is the detector dynamic resistance at the bias point. The \Eqref{SvRSJ} is valid both for a dc  and an ac biased TES \cite{SM}.

A more general derivation of the spectral density of the voltage fluctuation for a resistively shunted Josephson contact, with arbitrary current-phase relation $I(\varphi)$, has been derived by Nagaev \cite{Nag88}. The calculation is done for a junction biased at a  current $I>I_c$ in the region where the quasiparticle distribution function is close to equilibrium and the correlation function of the current is determined by the JN formula. The spectral density $S_V(\omega)$ at frequency $\omega<\omega_J$  is equal to
\begin{equation}\label{SvKN}
S_V(\omega)=R_d^2\left[1-\frac{\hat{V}}{2R_d^2}\frac{\partial R_d}{\partial I}\right]S_I(\omega).
\end{equation}
The two terms in the bracket describe the noise of the thermal fluctuations at equilibrium and should be experimentally compared with \Eqref{SvRSJ} derived for an RSJ with sinusoidal current-phase relation. In the derivation of   \Eqref{SvRSJ} and \Eqref{SvKN},  the system is considered to be in thermodynamic equilibrium and the fluctuation-dissipation theorem applies for the thermally excited quasiparticles \cite{RogScal1974} so  that $S_I(\omega)=4k_BT/\operatorname{Re}(Z_{tes}(\omega))$, i.e. the TES JN is treated as equilibrium thermal noise with the non-linearity  given by the Josephson impedance $Z_{mm}$.

The TES electrical noise can be calculated rather precisely from \Eqref{SvRSJ} and \Eqref{SvKN} at each bias point after estimating the detector electro-thermal parameters. This is done by evaluating the TES complex impedance $Z_{tes}(\omega,I_0,T_0)$ using  the standard technique of measuring the detector  current response to a voltage excitation at a given frequency $\omega$. The formalism for a dc and ac biased TES has been described in 
\cite{Lind2004} and \cite{Taralli2019}, respectively. 
The classical expression for $Z_{tes}(\omega)$ \cite{Lind2004} has been modified in \cite{Kozorezov11} to include the intrinsic reactance predicted by the RSJ model. The details are reviewed in  \cite{SM}.

The theoretical predictions described above can be verified experimentally.
By fitting the complex impedance curves taken along the transition, as described in \cite{SM}, we estimate the TES linear parameters $\alpha$, $\beta$ and the detector time constants for  many devices and bias frequencies. From $\beta$ and $R_d=R(1+\beta)$, we  calculate for each bias point the total TES voltage spectral density as in \Eqref{SvRSJ}.

\noindent  Evaluating the JN from \Eqref{SvKN} requires the derivative of the dynamic resistance with respect to the current, which cannot be accurately obtained from the $I-V$ characteristic. 

Our technique of measuring $Z_{tes}$ provides a straightforward estimation of $R_{d}$ and  $\partial R_d/\partial I$. 
We use the fact that $\partial R_d/\partial I=-R^3_d \partial^2 I/\partial V^2$, and that the TES current response  to a small and slow modulation $V_{m}\cos\omega t$ of the voltage bias, with $V_{m}<< V_{ac}$ and $\omega << \omega_b$, can be expanded as
\begin{equation}\label{Imod}
I(V,t)=I(V) + \frac{\partial I}{\partial V}V_{m}\cos\omega t +\frac{1}{2}\frac{\partial^2I}{\partial V^2}V^2_{m}\cos^2\omega t + ...
\end{equation}
The first harmonic is proportional to the inverse of the dynamic resistance $\partial I/\partial V=1/R_d$, while the second harmonic gives the term $\partial^2 I/\partial V^2$, which is needed to estimate the noise as in \Eqref{SvKN}. When measuring the intrinsic complex impedance curves, we fit the sinusoidal response up to the third order, to capture the harmonic distortions generated by the Josephson non-linearity as a function of bias point. An example of the fit of the response to a sinusoidal  excitation is shown in \cite{SM}. The coefficient of the first and second order term of the fit is used to accurately estimate the noise from the KN derivation. 
     
We measured the noise for many pixels along the TES resistive transitions. A typical data set is shown in \figref{fig:SvvsRRn} for a $80\times 40 \, \mu\mathrm{m}$ TES  biased at 2.5 MHz. 
The voltage noise, measured  in the kHz bandwidth, normalised to the thermal noise of resistance $R=\operatorname{Re}(Z_{tes}(\omega))$, is plotted as a function of $R/R_N$, along with the different noise  contributions. The noise shows the expected oscillating  behaviour from the Josephson effects being strongly correlated with the TES dynamic resistance. The first main result is that the NEJN dramatically underestimates the noise, in particular at low $R/R_N$ values. The total observed noise is consistent with the prediction from  \Eqref{SvRSJ} or  \Eqref{SvKN}, within the $1\sigma$ experimental uncertainties indicated by the shaded area around the lines. This result is in agreement with the fluctuation-dissipation theorem generalized for a non-linear system in thermal equilibrium.
The major contribution to the error bars derives from the uncertainties in the estimation of the bias circuit parameters, known within $10\%$, which propagate in the calculation of $V_{tes}$, $P_{tes}$, $G_{bath}$, $\alpha$ and $\beta$.   
The second important result is that both the RSJ model and the  derivation from KN explain equally well the detector noise. This implies that the  sinusoidal current-phase relation assumed in the RSJ model is a good approximation for the detectors and bias conditions discussed here. 
This is expected when a TES is treated as an SNS long diffusive weak-link  with Thouless energy $E_{TH}=\hbar D/L^2$, where  $D\simeq 0.018 \, \mathrm{m}^2/\mathrm{s}$ is  the diffusion coefficient for Au and $L=80-120 \, \mu \mathrm{m}$ the TES length, satisfying  the condition $E_{TH}\ll eV\ll k_BT$. 
The contribution of the ITFN noise for this particular design is typically small, and accounts for about $20\%$ of the total noise. The ITFN  is estimated from the Wiedemann-Franz law with $G_{tes}=L_0T/R_{\square}$, where $L_0$ is the Lorenz number \cite{ITFNnote}.
Some  excess noise is observed at $R/R_N>0.4$ for $T_{b}=55$ and $75\,\mathrm{mK}$, and it will be discussed later below.
\begin{figure}[H]
\center
\includegraphics[width=8.3cm]{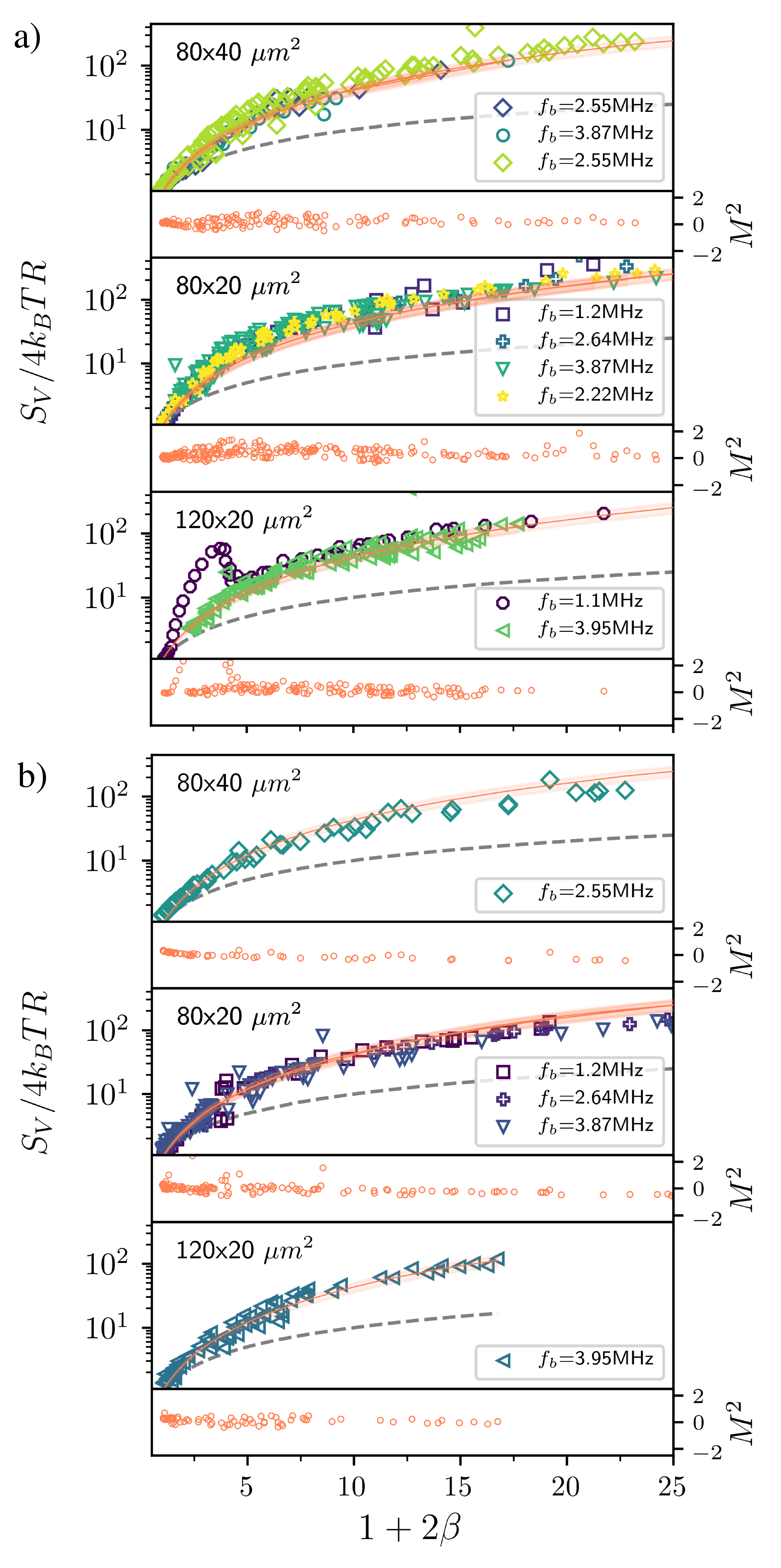}
\caption{\label{fig:Svvsbeta_all} Voltage noise, after ITFN subtraction, for three different TES design and several  bias frequencies, measured at   $T_{bath}=53 \, \mathrm{mK}$(a) and $T_{bath}=90 \, \mathrm{mK}$(b), respectively. The dashed line shows the noise estimation from NEJN. The red solid line is the prediction from the RSJ noise, with the shaded area indicating the 1$\sigma$ uncertainty on the calculated noise. The red open circle is the residual  $M^2$ calculated with respect to the RSJ model for all the measured pixels.}
\end{figure}
 To our knowledge, this is for the first time that the general derivation proposed by Nagaev \cite{Nag88} has been experimentally compared against the RSJ prediction. \Eqref{SvKN} was used before \cite{Lhotel2007} to explain the shot noise in the coherent regime of long diffusive SNS junctions at low bias voltage, $eV<E_{TH}$.   
In \figref{fig:Svvsbeta_all}, we summarise the noise measurement done for many other pixels with the three different TES geometries, biased at frequencies from 1.1 to 3.95 MHz. We plot  the voltage noise as a  function of the factor $1+2\beta$ (derived for the NEJN \cite{Irwin2006}), after subtracting the ITFN contribution. The solid red line is the expected  RSJ noise with  the shaded area highlighting the 1$\sigma$ uncertainties. The dashed line is the calculated NEJN. 
As shown by the small residual $M^2$ calculated using the RSJ model, the TES JN  is in very good agreement with the RSJ prediction over a large range of $\beta$ and for the high and low TES power regime ($T_{bath}=53$ and $90\, \mathrm{mK}$, respectively). The measured and expected noise is generally  independent on the bias frequency. 
Excess noise is observed at low value of $\beta$ ($\lesssim 3$)  with some pixels (for example  the $120\times 20 \, \mu\mathrm{m}^2$ biased at $1.1\, \mathrm{MHz}$, and few $80\times 20 \, \mu\mathrm{m}^2$ pixels at a $T_{bath}=53 \, \mathrm{mK}$ [\figref{fig:Svvsbeta_all}(a)]. 
\figref{fig:noisebumps} shows the details of  the observed excess noise for two devices as a function of $R/R_N$ and $T_{bath}$. 
\begin{figure}[h]
\center
\includegraphics[width=7.6cm]{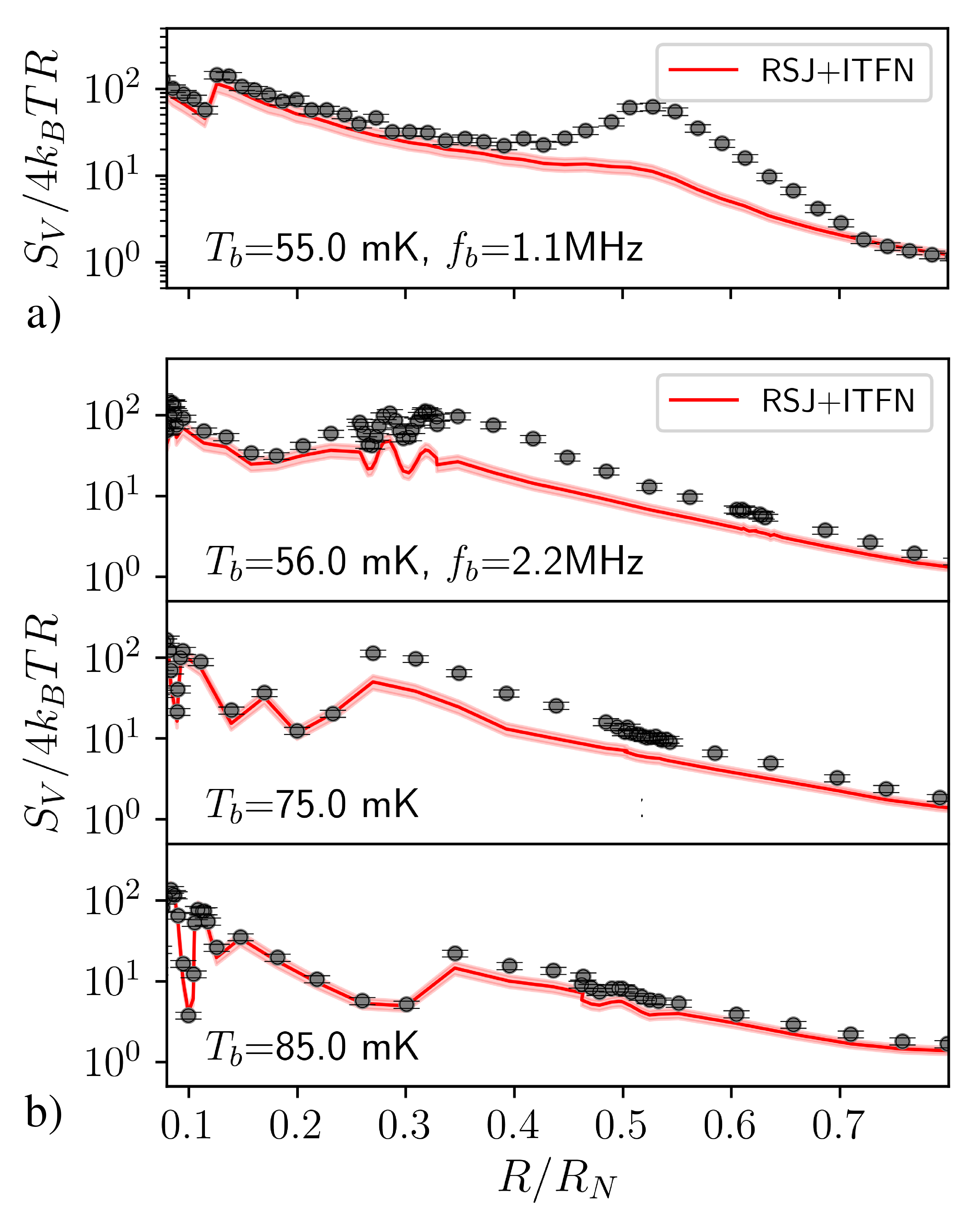}
\caption{\label{fig:noisebumps} Measured voltage noise at $T_{bath}=55.0 \, \mathrm{mK}$ of two $120\times 20 \, \mu\mathrm{m}$ TES as a function of bias point $R/R_N$ biased respectively at (a) 1.1  and (b) 2.2 MHz. The latter was measured at (from top to bottom) $T_{bath}= 56.0,\, 75.0$ and $85.0\, \mathrm{mK}$. The red curve is the expected RSJ and ITFN noise. }
\end{figure}
This noise is reduced when operating the device close to $T_c$, i.e. at lower TES  current \cite{SM}.
We believe this noise is of a different nature than the noise discussed above, which depends on the TES $R,\, \beta$ and $T$ and not on the current (\Eqref{SvRSJ}). It is likely due to a non-uniform current and temperature distribution inside the TES caused by the presence of normal metal structures (the stems connecting the absorber \cite{SM}) in the current path. The effect is minimized when operating the device close to $T_c$. From a recent investigation it seems  possible to eliminate the excess noise bumps by optimizing the size and  the position of the absorber-TES stems. A full study is under way.

In conclusion, we have extensively characterized at MHz bias the voltage fluctuations  of TES microcalorimeters  with three different design and $R_N$. When subtracting the expected ITFN noise, the residual noise in the JN bandwidth can be well explained, over a large range of experimental parameters ($R_N,V_{tes},f_{b},T_{bath}$), from the theory of noise in Josepshon weak-links, following the RSJ model or  the more general derivation from KN \cite{KogNag88}. The noise is consistent with the equilibrium thermal noise and is  enhanced by the non-linear response of the weak-link.  

\begin{acknowledgments}
This work is partly funded by European Space Agency
(ESA) under ESA CTP contract ITT AO/1-7947/14/NL/BW, by the European Union’s Horizon 2020 Program under the
AHEAD project (Grant Agreement Number 654215) and is part of the research program Athena (project number 184.034.002), which is (partly) financed by the Dutch Research Council (NWO).
\end{acknowledgments}

\widetext
\clearpage
\begin{center}
\textbf{\large Supplemental Materials: Voltage Fluctuations in ac Biased Superconducting Transition-Edge Sensors.}

\end{center}
\setcounter{equation}{0}
\setcounter{figure}{0}
\setcounter{table}{0}
\setcounter{page}{1}
\setcounter{section}{0}
\makeatletter
\renewcommand{\theequation}{S\arabic{equation}}
\renewcommand{\thefigure}{S\arabic{figure}}
\renewcommand{\thetable}{S\arabic{table}}
\renewcommand{\thesection}{S-\Roman{section}}

\section{The electro-thermal circuit and the derivation of the  complex impedance and the noise  of a TES behaving as a weak-link}
\begin{figure}[ht]
\center
\includegraphics[width=17.5cm]{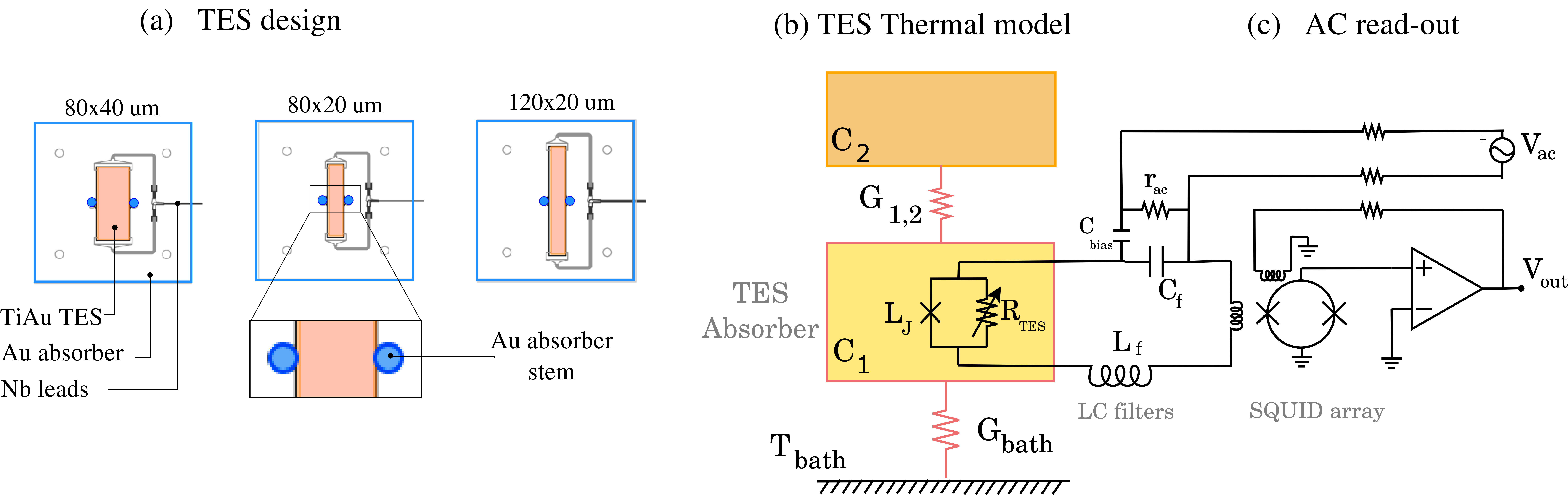}
\caption{\label{fig:ETscheme}. (a) Schematic diagram of the Ti/Au TESs studied in this work. Their geometry is, respectively, $80\times 40$, $80\times 20$ and $120\times 20\, \mu \mathrm{m}^2$, which correspond to an aspect ratio $L/W = $2,4 and 6. The TESs are connected to a $2.3\,\mu\mathrm{m}$ thick, $240\times 240 \, \mu\mathrm{m}^2$, Au absorber via the two central stems. The stems are in Au and have a diameter of $10\,\mu\mathrm{m}$. (b) a two-body thermal model of the TES calorimeter. (c) The MHz bias and read-out electrical circuit used     in the experiment. The detector is ac voltage biased in the superconducting transition, by means of an  high-{\it Q} {\it LC} resonator at frequency from 1 up to 5 MHz. The TES current is measured by an array of  SQUIDs amplifier.}
\end{figure}

We have characterized many Ti/Au TES microcalorimeters at different bias frequencies. The TESs have a critical temperature $T_\mathrm{c}\sim 90\, \mathrm{mK}$,  normal sheet resistance $R_{\square}=26\, \mathrm{m}\Omega/ \square$ and geometry $L\times W$ = $80\times 40$, $80\times 20$ and $120\times 20\, \mu \mathrm{m}^2$. 
A schematic diagram of the measured devices  is shown in \figref{fig:ETscheme}(a).
The normal resistance of the three TES design is, $R_N=52,104$ and $156\,\mathrm{m}\Omega$ and the measured thermal conductance-to-bath is $G_{bath}= 110,\,80$ and $110 \, \mathrm{pW/K}$, respectively. The TESs are thermally coupled to a $2.3\,\mu\mathrm{m}$ thick, $240\times 240\, \mu\mathrm{m}^2$ large Au absorbers by means of two Au stems, $10\,\mu\mathrm{m}$ in diameter, in the middle of the TES bilayer. The detectors are suspended on $0.5\,\mu\mathrm{m}$ SiN membranes. More details on the electro-thermal properties of these devices can be found in \cite{deWit_2020}.    
The TES-absorber system is thermally described with a two-body model schematically reported  in \figref{fig:ETscheme}(b). In line with the results from  \cite{Wake2019}, we believe the TES bilayer itself has a limited thermal conductance $G_{12}$ and $C_2$ is a fraction of the TES heat capacitance $C_{tes}$. $C_1=C_{tes}+C_{abs}$, where $C_{abs}$ is the heat capacitance of the absorber. For our devices $C_{abs}\gg C_{tes}-C_{2}>C_{2}$.

The read-out scheme is shown in \figref{fig:ETscheme}(c).  The detector is ac voltage biased in the superconducting transition, by means of an  high-{\it Q} {\it LC} resonator at frequency from 1 up to 5 MHz, and works as amplitude modulator. 
The MHz signal is amplified by an array of Superconducting QUantum Interference  Devices and demodulated at room temperature. After demodulation, the signal can be treated as in the dc bias case.  

The linearized system of the electrical and thermal differential equations are reported here in the frequency space and matrix form:   
\begin{equation}
\label{Mrsi}
\underbrace{
\begin{pmatrix}
z_{th}(\omega)+R(1+\beta) + dZ_{wl} & \mathcal{L}_0G_{bath}/I & 0 \\
-(2+\beta)IR-2IdZ_{wl}& G_{bath}(1-\mathcal{L}_0)+G_{12}+i\omega C_1 & -G_{12} \\
0 & -G_{12}  & +G_{12}+i\omega C_2
\end{pmatrix}}_{\mathcal{M}}\begin{pmatrix} \Delta I(\omega) \\ \Delta T_1(\omega)\\ \Delta T_2(\omega) \end{pmatrix}= \begin{pmatrix} v_{jn}\\ p_{ph}-Iv_{jn}\\ p_{2}\end{pmatrix}=\sqrt{S_{vp}},
\end{equation}
where $z_{th}(\omega)$ is the Thevenin equivalent of the bias circuit in \figref{fig:ETscheme},  $\mathcal{L}_0=I^2R\alpha/G_{bath}$  is the zero frequency the electrothermal feedback loop gain \cite{IrwinHilton,Lind2004} and $dZ_{wl}$ is the correction term to the TES impedance from the RSJ model as it is derived below. 
The noise terms, $\sqrt{S_{vp}}$ on the right-hand side of the equation are the detector  voltage and power noise sources described in the main text, with: $v_{jn}=\sqrt{S_{V,model}}$ the Johnson-Nyquist noise term, the main topic of our manuscript, $p_{ph}=\sqrt{S_{ph}}$ the phonon noise and $p_2=\sqrt{S_{itfn}}$ the ITFN noise. The total TES current noise, plotted in Fig.(1) of the main text, is obtained from  $i_{n,tes}=\sqrt{(\mathcal{MM^*})^{-1} S_{vp}}$. 

\begin{figure}
\center
\includegraphics[width=17.2cm]{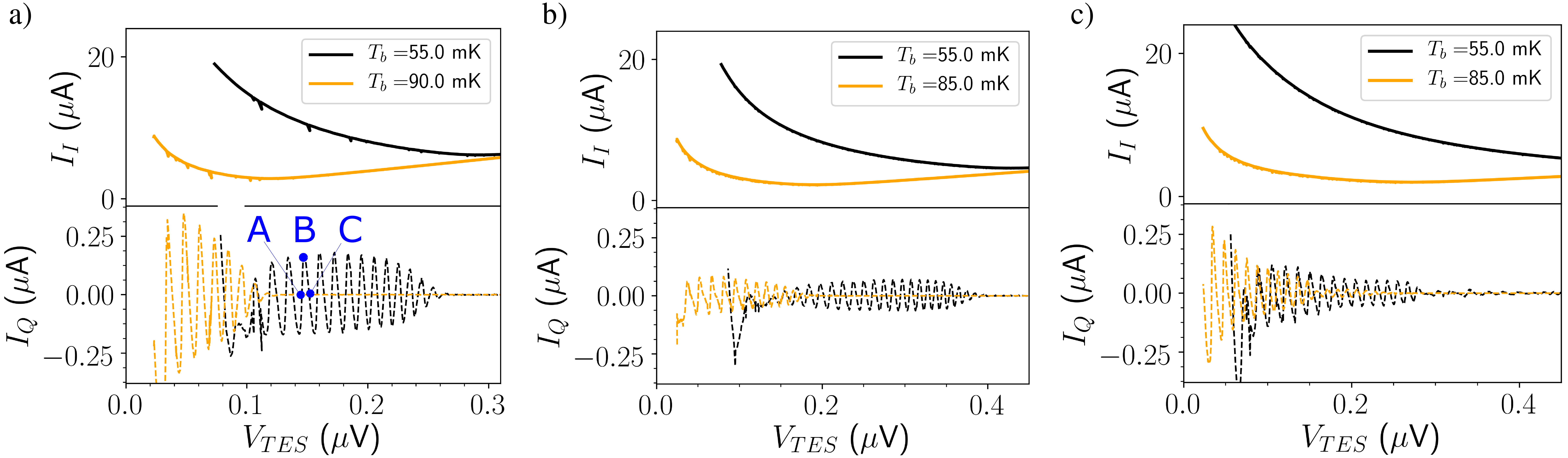}
\caption{\label{fig:ivs}In-phase, $I_I$, and quadrature. $I_Q$, TES current vs  voltage, for the three TES geometries, $80\times 40$ (a), $80\times 20$ (b) and $120\times 20$ (c) $\mu \mathrm{m}^2$,  all biased  at $f_{b}=2.5\, \mathrm{MHz}$. The $I_I,I_Q-V$ curves at $T_{bath}=\,55\, \mathrm{mK}$ (black lines), and  90 mK(a) and 85 mK(b,c) (orange lines) are shown. In (a), the A,B and C are the bias points discussed in \figref{fig:zwl} and in the related text. In colors on line.}
\end{figure}  

In \figref{fig:ivs}, we show the in-phase (solid line), $I_I$, and  quadrature (dashed line), $I_Q$,  for  the three TES microcalorimeter geometries. They represent respectively the {\it quasiparticles} and the {\it supercurrent} as described by the Josephson equations. The IV curves were  measured at $T_{bath}=55\,\mathrm{mK}$ (black) and $85,90\,\mathrm{mK}$ (orange), with the detectors biased at $\sim 2.5\, \mathrm{MHz}$. 
The period and amplitude of oscillation of the supercurrent as a function of the TES voltage can be described within the theoretical framework  based on the resistively shunted junction model (RSJ) \cite{Kozorezov11}, extended  to calculate the stationary non-linear response of a TES to a large ac bias current in the presence of noise \cite{Gottardi_APL14,Kirsch2020}.
The TES quiescent temperature $T$ is calculated from the $I_I-V$ characteristics and  the detector power low, $P=I_IV=k(T^n-T_{bath}^n)$, where $k=G_{bath}/n(T^{n-1})$. $G_{bath}$ is estimated by taking many $I-V$ curves at different bath temperature $T_{bath}$ and by fitting the $P-T_{bath}$ curve \cite{deWit_2020}.

In the main text, we have shown that the  voltage power spectral density of the  fluctuation $S_V(\omega)$ is the results of parametric conversion of fluctuations mixed  with the Josephson oscillations and its harmonics. 
Within the RSJ model, the low frequency power spectral density of the voltage fluctuations across the weak-link, averaged over the period of oscillation $2\pi/\omega_J$, is 
\begin{equation}\label{SvRSJx}
S_V(\omega)=R_d^2\left[1+\frac{1}{2x^2}\right]S_I(\omega),
\end{equation}
\noindent where $R_d=\partial V/\partial I=R(1+\beta)$ is the detector dynamic resistance at the bias point, and  $x=I/I_c(T)$, with $I$ and $I_c(T)$ the TES current and Josephson critical current respectively. 
To derive \Eqref{SvRSJx} we have used the following relations, which are valid for a RSJ with thermal energy $k_BT$ larger than 
$\hbar\omega_J$ and for $\omega\ll\omega_J$  \cite{KoganBook,LikSem72}:
$\left| Z_{00}(\omega)\right|=R_d$,  $\left| Z_{01}(\omega)\right|=\left| Z_{0-1}(\omega)\right|=\frac{1}{2x}\left| Z_{00}\right|$ and $S_I(\omega_J)=S_I(\omega)$.
We have not considered quantum correction since, for a typical TES,  $k_BT\gg eV$.  The \Eqref{SvRSJx} is valid both for a dc  and ac biased TES.
We can approximate \Eqref{SvRSJx} using the solution of the Josephson equations \cite{AslLar1969} in  the RSJ model, valid for $I>I_c$.  The current-voltage characteristic, averaged over a period of oscillations, $2\pi/\omega_J$, becomes then $\hat{V}=I_cR_N\sqrt{x^2-1}$,with $R_N$ the resistance of the weak-link in the normal state. From the definition of the dynamic resistance, and with $R=\operatorname{Re}(Z_{tes})$, it can be shown \cite{KoganBook,Kozo12} that  $1/2x^2=1/2\left[1-(R_N/R)^2/(1+\beta)^2\right]$, which, from \Eqref{SvRSJx}, leads to 
Eq.~(1) of the main text.  

\begin{figure}
\center
\includegraphics[width=15cm]{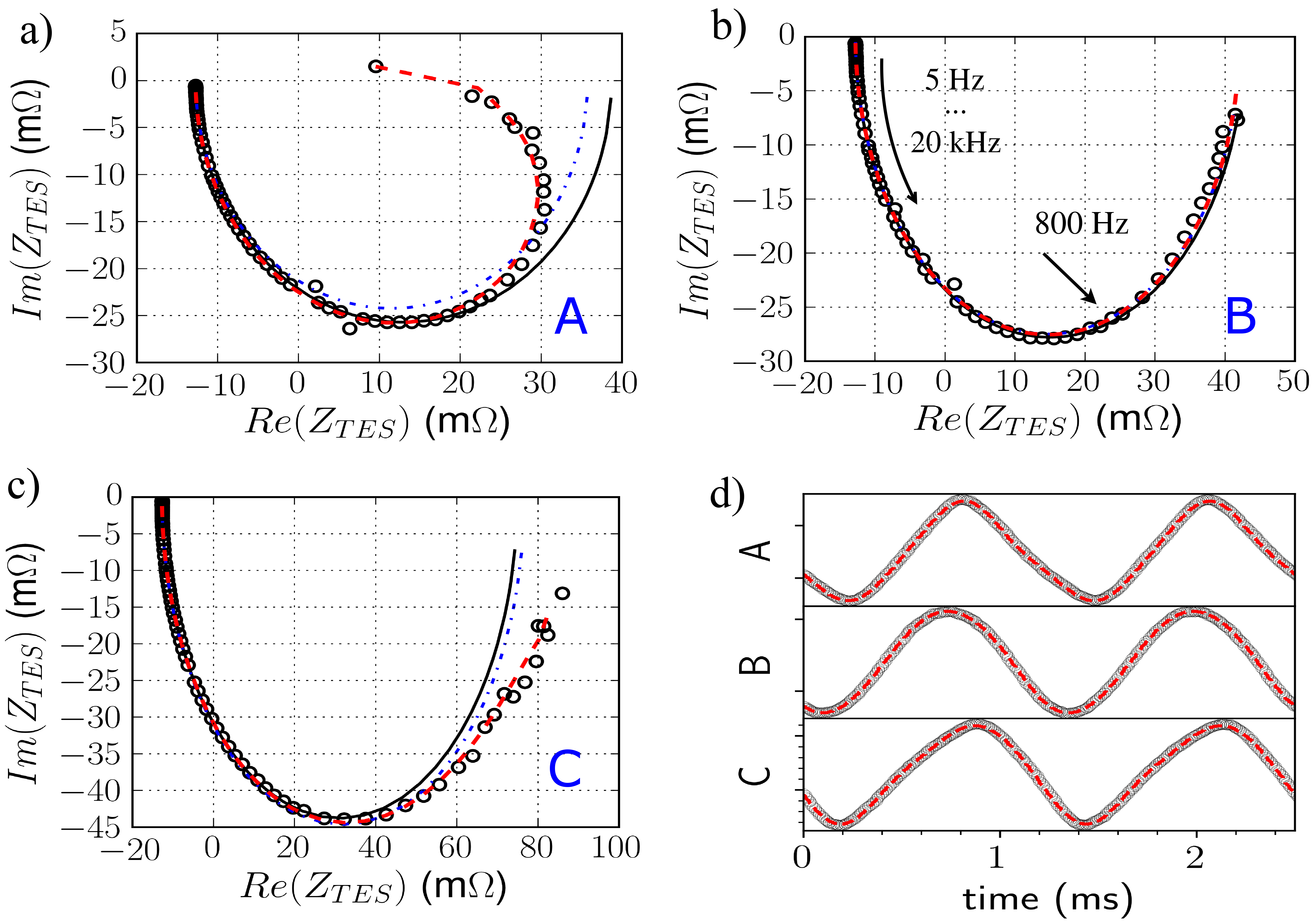}
\caption{\label{fig:zwl} (a,b,c) Complex impedance $Z_{tes}(\omega))$ taken at the A,B and C bias points ($R/R_N\sim 30\%$) shown in \figref{fig:ivs}(a). The solid black and dash-dotted blue lines are the fit to a single and two-body thermal model, respectively. The red dashed line is the best fit with a two-body and the RSJ model. (d) Example of the detector response to a sinusoidal small excitation at $800\, \mathrm{Hz}$. The red solid line is the fit to 
Eq.~(3) in the main text. The curves have been taken with the TES operating at a bath temperature $T_{bath}=55\, \mathrm{mK}$}
\end{figure}

The classical expression for the TES complex impedance $Z_{tes}(\omega)$ \cite{Lind2004,IrwinHilton} has been modified by including the intrinsic reactance predicted by the RSJ model. From the results in \cite{Kozorezov11} and the discussion reported in \cite{Zorin81,Vyst74}, the expression for $Z_{tes}(\omega)$, for a TES ac biased at a frequency $\omega_b$, becomes:
\begin{equation}\label{Ztesac}
Z_{tes}(\omega)=Z^0_{tes}(\omega)+dZ_{wl}(\omega)=Z^0_{tes}(\omega)+\frac{1}{1-\mathcal{L}_0}\frac{Z_{00}(\omega+\omega_b)-Z_{00}(\omega_b)}{1-i\omega\tau_{eff}},
\end{equation}   
\noindent where $Z^0_{tes}(\omega)$ is the classical TES  impedance, which, for the two-body model described before, takes the form 
\begin{equation}
Z^0_{tes}(\omega)=R\left[1+\beta+\frac{(2+\beta)\mathcal{L}_0}{i\omega\tau_{01}-\mathcal{L}_0+1-G_{12}/(G_1+G_{12})}\frac{1}{1+i\omega\tau_{02}}\right]
\end{equation} 
with $\tau_{01}=C_1/(G_{bath}+G_{12})$ and $\tau_{02}=C_2/G_{12}$ \cite{Maasilta12}. The element $Z_{00}(\omega)$ is derived analytically in \cite{Kozorezov11} using the solution for the non-linear impedance of an ac driven Josephson junction \cite{Coffey00}. 
An example of the intrinsic complex impedance of the same device is shown in \figref{fig:zwl}(a,b,c), respectively taken at the  three different bias points (A,B,C) indicated in the $I_{Q}-V$ (at  $R/R_N\sim 0.3$).

The $Z_{tes}(\omega)$ data points are the results of a small sinusoidal excitation between $5\,$Hz and $20\,\mathrm{kHz}$ \cite{Taralli2019} away from the bias frequency of $2.5\,\mathrm{MHz}$.   
The solid black and dash-dotted blue lines are the fit of the data, with respectively a single-body and a two-body thermal model \cite{Wake2019,Maasilta12}. The dashed red line is the best fit with a two-body thermal model and  including \Eqref{Ztesac}.

 \noindent As predicted in \cite{Kozorezov11}, we observe inward and outward inflexions of the complex impedance at high excitation frequency, which cannot be explained by a thermal effect, and are the results of the RSJ intrinsic reactance. 

As an example, we show in \figref{fig:zwl}(d) the detector response to a sinusoidal small voltage $V_m\ll V_{ac}$ excitation at $800\, \mathrm{Hz}$, when it is biased at the A,B and C  points shown in \figref{fig:ivs}(a). The red solid line is the fit to 
Eq.~(3) in the main text. The  800 Hz frequency data point  has been chosen for the example because it is typically where the response of the TES to the Johnson noise is at its maximum, for this specific bias point. The fit to 
Eq.~(3) was done for all the frequency points, from 5 Hz up to about 20 kHz, shown in the complex impedance plots \figref{fig:zwl}(a,b,c).

\newpage

%


%

\end{document}